# Data Center Control Against Sub-Synchronous Resonance: A Data-Driven Approach

Grant Ruan[1*], Marija D. Ilic[1], Le Xie[2]
*CIGRE Member ID: 620250370
*Postdoc, Young Professional with 4 years
[1]Massachusetts Institute of Technology, [2]Harvard University
gruan@mit.edu, ilic@mit.edu, xie@seas.harvard.edu

## SUMMARY

As the backbone of modern digital infrastructure, data centers host a variety of essential services such as cloud computing and artificial intelligence. Electric grid operators, however, have limited experience and knowledge of the reliability risks of data center interconnection due to their unique operational characteristics. An emerging concern is the sub-synchronous resonance (SSR) which refer to unexpected voltage/current oscillations at typical frequencies below 60/50 Hz. It remains unknown whether and how the interactions between data centers and the grid may trigger resonances, equipment damages, and even cascading failures.

In this paper, we focus on grid-connected data centers that draw electricity from the grid through power factor correction (PFC) converters. We conduct two-tone frequency sweep to investigate the data centers' impedance characteristics, i.e. magnitude and phase angle variations over frequencies, and showcase their deep dependence on compute workloads. The impedance modeling provides a direct approach to evaluating SSR risks and enable a cooperative mechanism to alarm and avoid resonance-prone situations. Building upon the impedance, a data-driven preventive controller is then established to raise early warnings of risky operation and suggest flexible workload management according to the given grid conditions. Through case study, we demonstrate how to use impedance to understand the unexpected interactions. Data-driven impedance is validated to show decent performance in capturing the unique impedance dips and tracking the impedance variations across a range of workload scenarios. The early warning and preventive control approaches are further effective to improve the safety margins with minimal workload rescheduling. The key findings of this work will provide valuable insights for grid operators and data center managers, and support preparation for future scenarios involving large-scale data center integration.

# 1. INTRODUCTION

Sub-synchronous resonance (SSR) has emerged as a major challenge to the dynamic security of modern power systems. SSR is formally defined as an oscillation phenomenon with large-scale, undamped energy exchanges repeating at frequencies below 60/50 Hz (typically 5~55 Hz). Concerns about SSR have steadily grown in the past decades due to its unique feature of self-excitation and negative damping; in other words, stability collapse may happen even without forced injections.

While many SSR incidents were tied to wind generation[1], data centers have emerged as a potential new contributor. Such demand-side risk is no longer hypothetical as oscillations and low-frequency harmonics are increasingly observed in real-world data centers. For example, an 11 Hz resonance was found in an area near several Meta data centers in 2017[2], with current measurements fluctuating all the way up to utility substations. During the summer of 2024, another 14.7 Hz oscillation emerged from a cluster of data centers in Virginia[3] and caused a two-hour periodic voltage sag (see *Fig. 1*). This resonance was confirmed to originate from the input stage of uninterruptible power supply. Technically, both events are attributed to undesirable controller interactions, which differ from the past experience of wind turbines. Therefore, it is urgent to establish a thorough understanding of these new patterns.

Data centers are emerging large loads that raise new risks to grid reliability[4]. Several community groups are pushing forward the task forces to understand their unique dynamics, including NERC[5] and ERCOT[6]. However, most of the current attentions are focused on data centers' energy mangement[7] or converter designs[8]. Many works still rely on linearization and small signal analysis, leaving large-signal modelling relatively limited in the literature. Research on data center dynamics and the potential risks of SSR are still in early stage.

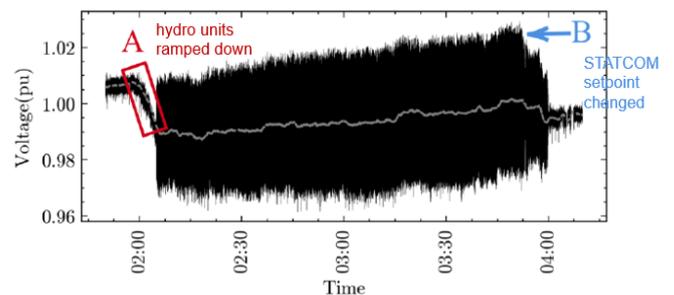

*Fig. 1. A two-hour sub-synchronous resonance measured at a Virginia substation. Source: [3].*

This paper is motivated to study grid-connected data centers and provide control-theoretical insights into how data centers may cause SSR incidents and destabilize the grid. The efforts will contribute to relieving the reliability concerns of data centers and then accelerating the interconnection process of large-scale digital facilities.

The major contributions of this paper are summarized as follows:

- A novel framework is proposed to coordinate data center managers with grid operators at the preventive stage of SSR mitigation. Impedance adaptation is formalized as a measure in this framework to avoid degraded damping and enhance the grid compatibility of data centers.
- A workload-dependent impedance model is established to characterize the dynamical responses of data centers across multiple frequency points. We apply large-signal models and frequency sweep to consider all nonlinear power electronic dynamics inside the power factor correction converters. The unique characteristics of magnitude dips are well captured.
- A preventive controller is developed to maximize the worst-case safety margin and mitigate SSR by flexibly optimizing the workload level. We find the basin of safety margins for moderate-to-high workloads and validate that the common deloading strategy may cause stability degradation.

# 2. FRAMEWORK

## 2.1 Grid-Connected Data Centers

The main goal of this paper is to study SSR by deconstructing the rich interactions between data centers and the electric grid. In specific, we focus on a grid-connected data center that is operating to handle diverse compute workloads. We consider a typical interconnected system as shown in *Fig. 2*, and the impacts of data centers on local grid can be isolated and analyzed in detail.

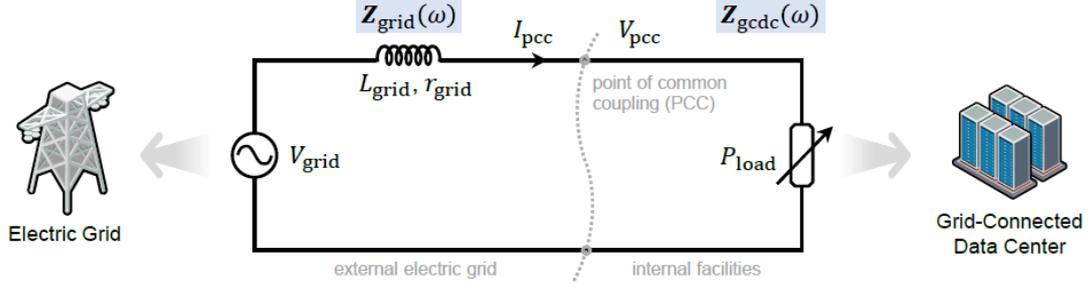

*Fig. 2. An interconnected system to study how data centers interact with the grid*

The interconnected system is separated into an external and an internal part, where we consider a Thevenin equivalent for the external grid (a voltage source plus a series inductance) and a constant power load for the data center (nonlinear, workload setpoints may vary). We account for harmonic distortion and don't assume perfect sinusoidal power source.

It is classical to express the grid-side impedance $\mathbf{Z}_{\text{grid}}$ and the data center impedance $\mathbf{Z}_{\text{gcdc}}$ as complex-valued functions of frequency in either a rectangular or a phasor form:

$$\mathbf{Z}_{\text{grid}}(\omega) = r_{\text{grid}} + jwL_{\text{grid}} = \sqrt{r_{\text{grid}}^2 + w^2 L_{\text{grid}}^2} \angle \tan^{-1}(wL_{\text{grid}}/r_{\text{grid}}) \quad (1)$$

$$\mathbf{Z}_{\text{gcdc}}(\omega) = \|\mathbf{Z}_{\text{gcdc}}(\omega)\| \angle \mathbf{Z}_{\text{gcdc}}(\omega) \quad (2)$$

Note that the above two equations are large-signal impedance models that can cover multiple operation conditions and the full spectrum of frequencies. The detailed representation of $\mathbf{Z}_{\text{gcdc}}$ will be presented in the next section.

## 2.2 Potential Sub-Synchronous Resonances

A primary concern about interconnected systems is the potential occurrence of SSR. Here, a control-theoretical viewpoint is applied to study the resonance mechanisms and illustrate how grid-connected data centers may excite oscillatory modes.

The interconnected system of Fig.1 can be cast as a **unity feedback system** (current-based, voltage feedback). A classical approach in resonance analysis is to break the loop at the ground node, deactivate the voltage source, and then evaluate the remaining open-loop gain as follows:

$$\mathbf{G}_{\text{ol}}(\omega) = \mathbf{Z}_{\text{grid}}^{-1}(\omega) \cdot \mathbf{Z}_{\text{gcdc}}(\omega) \quad (3)$$

It is stated in Nyquist stability theorem that the closed-loop performance is dictated by the associated open-loop gain. In this sense, open-loop analysis is sufficient (and also convenient) to provide critical details of the closed-loop interconnected system of interest.

Next, we move to the resonance status and resonance conditions.

Resonances relate to the critical status of sustained oscillations (stability boundary), which are characterized by unity gain and a half-cycle phase lag. This can be translated into:

**Sub-Synchronous Resonance Condition**: The SSR is detected to take place in an interconnected system, when $\exists w_{\text{ssr}} \in \{0, w_g\}$ such that either of $w = w_g \pm w_{\text{ssr}}$ satisfies the formula below.

$$\mathbf{Z}_{\text{grid}}^{-1}(\omega)\, \mathbf{Z}_{\text{gcdc}}(\omega) = -1 \quad (4)$$

The complex-valued criterion (4) might be inconvenient for direct evaluation, and a common alternative is to use Bode plots. By drawing impedance $\mathbf{Z}_{\text{grid}}$ and $\mathbf{Z}_{\text{gcdc}}$ in the same plot, a resonance can be detected if their magnitude curves intersect at a frequency with a phase lag of 180 degrees.

$$\|\mathbf{Z}_{\text{grid}}(\omega)\| = \|\mathbf{Z}_{\text{gcdc}}(\omega)\|; \quad \angle \mathbf{Z}_{\text{grid}}(\omega) - \angle \mathbf{Z}_{\text{gcdc}}(\omega) = 180° \quad (5)$$

## 2.3 Impedance Adaptation to Mitigate Resonance

**Impedance Adaptation** is proposed and established as a mitigation measure against SSR. The key idea is to proactively manage workloads in order to adapt the impedance of data centers and prevent their potentially adverse interactions with the grid.

The development of impedance adaptation requires two major steps: We first investigate the dependence between workloads and impedance characteristics, and then develop metrics to guide the workload management.

*1) Workload Dependence.* Data centers are known to be nonlinear loads and their V-I characteristic, impedance $\mathbf{Z}_{\text{gcdc}}$ both depend on the operating points. Workload is one of the few factors to specify an operation point. In this context, we express the workload dependence by extending the impedance notation to $\mathbf{Z}_{\text{gcdc}}(w|P_{\text{load}})$. The specific representation is deferred to the next section.

*2) Distance-Based Evaluation.* We instantiate the classical Nyquist distance concept by supplementing the workload dependence. This metric describes how close a stable system is to the stability boundary (4) and it takes account of both magnitude and phase simultaneously.

$$\text{dist}(w, P_{\text{load}}) = \left\| 1 + \mathbf{Z}_{\text{grid}}^{-1}(\omega)\, \mathbf{Z}_{\text{gcdc}}(\omega|P_{\text{load}}) \right\| \tag{6}$$

The proposed metric (6) is useful to decide impedance adaptation against SSR. The idea is to address a potential resonance frequency point by restricting the workload (typically deloading) so as to keep apart from the stability boundary. This physically translates to an increase in closed-loop gain and better damping characteristics.

The subsequent sections will supplement this mitigation measure with great details. In particular, Section 3 is focused on the impedance derivation of $\mathbf{Z}_{\text{gcdc}}(w|P_{\text{load}})$ through frequency sweep, and Section 4 generalizes the metric to account for worst-case robustness and run an optimization to identify the best level of workloads. All these additional details adhere to the proposed framework while instantiating it in concrete settings.

An effective impedance adaptation relies on the close cooperation between data centers and the grid. A cooperative mechanism is proposed and the specific process are elaborated in *Fig. 3*. As shown, the entire coordination is initiated by grid operators who are expected to share the information of (estimated, average or worst case) grid impedance for routine checks or known/predicted future events. To match the grid-side inputs, the data center will apply impedance adaptation to decide the appropriate level of workloads that prevents SSR risks. The actions will be executed once the next decision cycle or event takes place.

## 2.4 Data-Driven Solution for Impedance Adaptation

Our work is unique in the use of data-driven models and approaches. In *Fig. 3*, the data-driven components are all marked blue, which distinguish the existing analytical solutions. Note that learning-based modeling is preferred because of the decent balance between accuracy and computational speed; and high-fidelity optimization for hard non-convex problems.

Analytical computation of impedance spectrum is intensive (up to hours) and impractical for most online applications. Deep learning models are often more effective in online settings due to the workflow of offline training and online deployment.

With deep learning models embedded, many optimizations tend to become highly nonlinear, non-convex, and the gradient computation is less reliable. Beyasian optimization is a data-driven optimization method leveraging data acquisition to guide the search for optimal solutions.

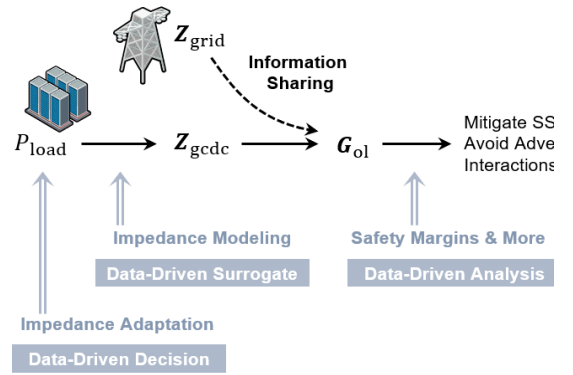

*Fig. 3. Cooperation process*

In the subsequent sections, we will demonstrate the details of data creation at the end of Section 3 and all the data-driven modeling in Section 4.

## 3. IMPEDANCE MODELING OF DATA CENTERS

### 3.1 Power Supply Architecture

This section intends to establish the detailed representation of the data center impedance $\boldsymbol{Z}_{\text{gcdc}}$. The fact that a system's impedance is intrinsically govern by the internal circuit configurations has motivated us to examine the power supply architecture within a mainstream data center.

Modern data centers configure their power chain as follows. The AC power drawn from the utility feeder is first stepped down by AC/AC transformers and converted to DC through an AC/DC rectifier. The DC power is then processed through DC/DC isolation and modulation, and finally delivered to onboard chips to execute compute workloads.

This power chain deploys controllers mainly at the AC/DC and DC/DC stages, following the classical scheme of hierarchical bandwidth allocation. In specific, the converters that are far from the terminal workloads are designed to exhibit slower dynamical responses and wider bandwidths. As a consequent, the AC/DC conversion stage becomes the global bottleneck of the transient process of data centers, and thus play a major role in shaping the impedance characteristics.

The mainstream converters for AC/DC conversion are the so-called power factor correction (PFC) converters, whose modeling details will be presented in the next subsection. For the purpose of this work, we simplify the AC/AC side details; we also neglect the ultra-fast dynamics of DC/DC stages and treat the DC side as a constant-power load. In contrast, we examine all the power electronic details in PFC converters, which are found to cause special impedance characteristics of magnitude dips. Our model provides sufficient impedance details so that SSR risks can be effectively assessed.

A typical architecture is presented below.

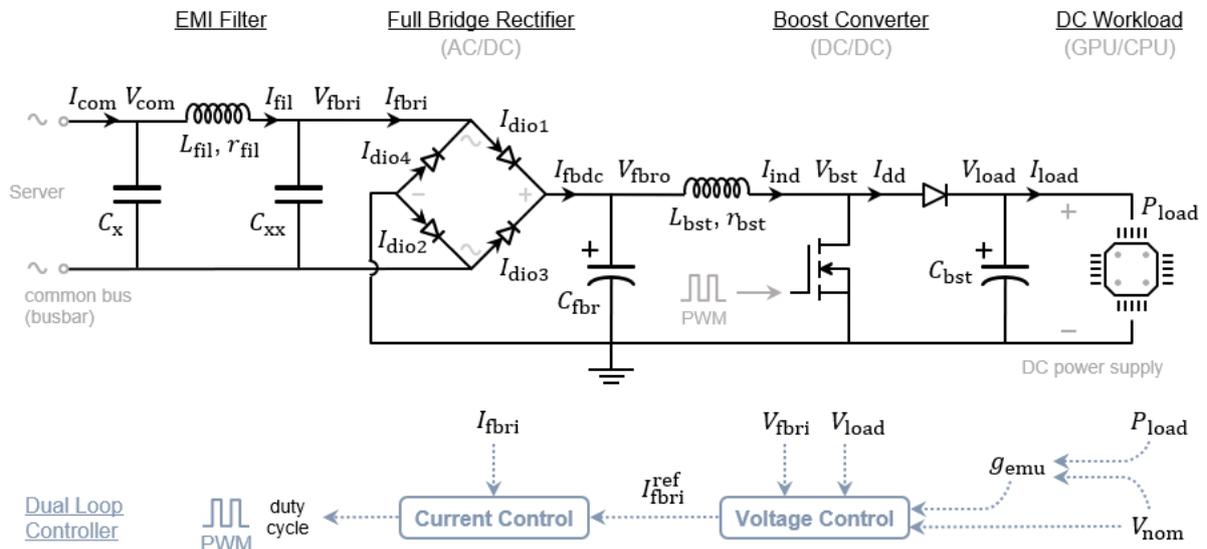

Fig. 4. Power factor correction converter topology and control.

### 3.2 Power Factor Correction Converters

PFC converters are advanced power electronic modules to align input currents and voltages in order to reduce the potential harmonics and reactive power injections. These converters are widely adopted in modern data centers as they can lower energy losses and contribute to compliance with grid codes at low cost. As shown in *Fig. 4. Power factor correction converter topology and control.*, a typical PFC converter includes an electromagnetic interference (EMI) filter, a full-bridge rectifier, and a boost circuit.

This subsection will set up the dynamical models of PFC converters. We start with one PFC converter whose circuit topology is detailed in *Fig. 4*. The input is the common bus voltage $V_{\text{com}}$ (with harmonics) and the workload $P_{\text{load}}$, and the output is the status of all state variables $\vec{x}_{\text{pfc}}$. A simple selector can be applied to extract the common bus current $I_{\text{com}}$ (with harmonics) from $\vec{x}_{\text{pfc}}$, formally:

$$\vec{x}_{\text{pfc}} = [V_{\text{com}}, I_{\text{fil}}, V_{\text{fbri}}, V_{\text{fbro}}, \cdots, \Sigma_v, \Sigma_i]^\top \tag{7}$$

$$I_{\text{com}} = I_{\text{fil}} = \vec{e}^\top \vec{x}_{\text{pfc}} \tag{8}$$

where $\vec{x}_{\text{pfc}}$ is represented by the capacitor voltages, inductance currents, and two integral terms from voltage/current control; $\vec{e} = [0, 1, 0, \cdots]^\top$ is a sparse vector for variable selection.

The operation of PFC converters can be fully captured by a high-dimensional ordinary differential equation. Below, we use $\text{ODE}(\cdot)$ to denote this aggregate dynamic of the entire circuit:

$$\frac{d}{dt}\vec{x}_{\text{pfc}} = \text{ODE}(\vec{x}_{\text{pfc}} | P_{\text{load}}, V_{\text{com}}) \tag{9}$$

The detailed formulations of $\text{ODE}(\cdot)$ cannot be fully expanded due to the space limit. It basically contains all the circuit elements' constitutive laws and the system Kirchohoff's voltage and current law. The diodes and MOSFET are configured with parasitic resistors to avoid stiff fluctuations. It should be noted that this is a highly nonlinear system involving high-frequency switching (discrete jumps), complicated diode dynamics (exponential-type), and control-induced cross-product terms. As a consequence, the AC-side waveforms may contain rich harmonics, and the DC side exhibits voltage and current fluctuations with small ripples.

PFC converters implement two PI controllers in an outer voltage loop and an inner current loop. Voltage control adjusts the current reference based on DC voltage fluctuations, while current control adjusts the PWM duty cycle to track the reference and the input voltage waveform. In closed loop, PFC converters implement the current waveform modulation for reduced harmonics and near-unity power factors.

### 3.3 Workload-Dependent Impedance Model

In this subsection, we examine how workload can influence impedance and formulate a workload-dependent impedance model to take account of their dependence. Workload is selected because of the availability of accurate measurement data and the ease of implementation without hardware upgrades. In fact, workload management system is widespread in data centers, so controlling workload is technically convenient. Note that the impedance adaptation can be realized rapidly through minimal software designs only.

Our primary focus is on how $\mathbf{Z}_{\text{gcdc}}$ is influenced by $P_{\text{load}}$ within the frequency range of $(0, 2w_g)$.

Accurate impedance modeling calls for a full consideration of the nonlinear characteristics and discrete switching behaviors. Therefore, we run precise circuit simulations combined with frequency sweeps to compute the PFC impedance across a range of operating points. In the next section, a data-driven surrogate is proposed for approximation. Note that impedance of multiple operating points can only be calculated by large-signal models because small-signal linearization is no longer valid.

*1) Dual Port Harmonic Analysis*: This is a necessary step to understand how to set up the frequency sweep. To see this, we form a dual-port network in *Fig. 5* to analyze the harmonic frequency of $w_{\text{hm}}$. The nonlinearity of PFC converters will induce cross-frequency couplings (intermodulation distortion) between $w_g$ and $w_{\text{hm}}$, and the impedance to compute $\mathbf{Z}_{\text{gcdc}}(w_{\text{hm}})$ is essentially counting the contributions from both frequencies.

The above idea neglects the harmonic coupling across other frequencies and concentrates on the dominant component of $w_{\text{hm}}$. This simplification avoids the use of impedance matrices (instead of a scalar) in the downstream stability assessment. In practice, this assumption is valid because real-world harmonic injections are always small.

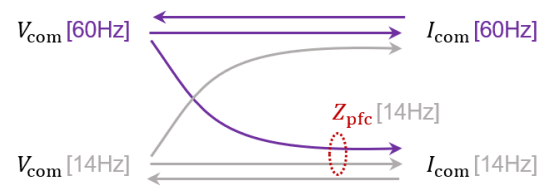

*Fig. 5. Dual-port harmonic analysis.*

*2) Two-Tone Frequency Sweep*: We apply a dual-sine sweep to evaluate the PFC impedance. This sweep is fundamentally different from the classical single-tone sweep that is no longer valid for nonlinear PFC converters. In specific, the following excitation source is chosen for the sweep:

$$V_{\text{com}}^{\text{sweep}} = V_{\text{com}} \cos(w_g t + \theta_0) + \Delta V_{\text{hm}} \cos(w_{\text{hm}} t) \tag{10}$$

where $\Delta V_{\text{hm}}$ is set as 5% of the $V_{\text{com}}$ magnitude; the initial phase difference $\theta_0$ has a limited impact ($< 3\%$) on the final estimation, so we omit scanning over it for brevity.

Next, we simulate the PFC converters using the source input $V_{\text{com}}^{\text{sweep}}$ to get the time-domain response of $I_{\text{com}}$. Without loss of generality, we assume the simulation starts from a steady state $\vec{x}_{\text{pfc}}^0$ and a sequence of length $K$ is collected accordingly.

$$\{\vec{x}_{\text{pfc}}\}_{\text{seq}} = \left\{ \vec{x}_{\text{pfc}}^0 + \int_0^{k\Delta t} \text{ODE}(\vec{x}_{\text{pfc}} | P_{\text{load}}, V_{\text{com}}) dt \right\}_{k=0}^{K-1} \tag{11}$$

$$\{I_{\text{com}}\}_{\text{seq}} = \{\vec{e}^\top \vec{x}_{\text{pfc}}\}_{\text{seq}} = \vec{e}^\top \{\vec{x}_{\text{pfc}}\}_{\text{seq}} \tag{12}$$

The calibration of $\vec{x}_{\text{pfc}}^0$ and $K$ are two critical details. It is straightforward to start with a proper initial state and approach $\vec{x}_{\text{pfc}}^0$ until the transients die out. The sequence length $K$ must be larger than the reciprocal of the expected frequency accuracy and further extended to ensure perfect periodicity and avoid spectrum leakage errors.

Through Fourier analysis, we are able to extract the harmonic component of $w_{\text{hm}}$, and estimate the impedance as the ratio of the harmonic voltage and current phasors. Note that the temporal resolution of circuit simulation is always precise enough to capture frequencies up to $2w_g$; the harmonic voltage is $\Delta V_{\text{hm}} \angle 0°$ in the computation.

$$I_{\text{hm,fft}}(w_{\text{hm}}) = \text{FFT}(\{I_{\text{com}}\}_{\text{seq}}, w_{\text{hm}}) \tag{13}$$

$$Z_{\text{pfc}}(w_{\text{hm}}) = \frac{\Delta V_{\text{hm}}}{\|I_{\text{hm,fft}}(w_{\text{hm}})\|} \angle \left(-I_{\text{hm,fft}}(w_{\text{hm}})\right) \tag{14}$$

*3) Impedance Data Generation*: The above process completes one estimation point for $Z_{\text{pfc}}$, and it should be repeated until fully covering the frequency range of $(0, 2w_g)$. We apply a classical grid search of the workload and frequencies to create an impedance dataset. This dataset does not exhaust all the operating conditions (infinite) and will be fitted in the next section using a learning-based model. In addition, the overall data center impedance is a combination of these PFC converters at the connection port, and parallel-connected aggregation is commonly used to build up these connections.

## 4. DATA-DRIVEN MITIGATION OF SUB-SYNCHRONOUS RESONANCE

### 4.1 Data-Driven Surrogate for Impedance

Precise impedance calculation is time-consuming and unsuitable for online applications. In this regard, data-driven models can provide a practical alternative to strike a balance between accuracy and computational efficiency. The key idea is to learn a direct mapping from workloads and harmonic frequencies to the associated impedance characteristics. Formally, the proposed data-driven surrogate is denoted by $\hat{Z}_{\text{gcdc}}(w|P_{\text{load}})$, with the following estimator:

$$\left[\text{Re}(\hat{Z}_{\text{gcdc}}), \text{Im}(\hat{Z}_{\text{gcdc}})\right]^\top = \text{NN}_\phi(w, P_{\text{load}}) \tag{15}$$

Technically, this is a standard regression task, and we apply the classical dense neural nets with the tanh activation function to ensure smoothness. The impedance dataset created in the previous subsection is partitioned into a training (80\%) and a test (20\%) set.

A special detail of this neural net is the impedance representation, i.e. either phasor or rectangular forms. Our simulation suggests that the rectangular form performs more stable and accurate. To see this, small errors in phasor representation can cause disproportionately large relative errors.

### 4.2 Safety Margins and Early Warnings

This subsection will assess the SSR safety in a given workload condition. The distance metric from Subsection 2.3 is just a starting point, and it will be used to measure the safety margins during a full frequency scan.

**SSR Safety Margin** is an indicator quantifying how far a system is away from the adverse conditions that would trigger SSR events. These margins can be determined using the distance metric from Subsection 2.3. The idea is to run a worst-case search to find out the minimum distance and the fragile frequency condition:

$$\min_{w \in [0, 2w_g]} \text{dist}(w, P_{\text{load}}) \Rightarrow M_{\text{ssr}}(P_{\text{load}}), w_{\text{vul}} \tag{16}$$

where $M_{\text{ssr}}$ denotes the safety margin and $w_{\text{vul}}$ denotes the most vulnerable frequency condition; we intentionally close $w \in (0, 2w_g)$ at the boundary points to ensure reachable minimizers.

Using the data-driven surrogates, the safety margin can be formulated as follows:

$$M_{\text{ssr}}(P_{\text{load}}) \approx \min_{w \in [0, 2w_g]} \| 1 + \mathbf{Z}_{\text{grid}}^{-1}(\omega) \hat{\mathbf{Z}}_{\text{gcdc}}(\omega | P_{\text{load}}) \| \tag{17}$$

Note that the above is a hybrid combination of analytical and neural net components, which is generally hard to solve. One critical challenge is the absence of reliable gradients, and therefore, we apply Bayesian optimization, a popular derivative-free solver with good probabilistic guarantees.

An absolute-value indicator like (17) is sometimes non-intuitive to understand and provides limited information about the performances. To improve interpretability, a normalized indicator $\bar{M}_{\text{ssr}}$ is built by taking the nominal distance of grid frequency as a comparison baseline. The formal definition is given below:

$$\bar{M}_{\text{ssr}}(P_{\text{load}}) = M_{\text{ssr}}(P_{\text{load}}) / \text{dist}(w_g, P_{\text{load}}) \tag{18}$$

**Early Warning** serves as a static assessment process that uses the above two indicators to identify potential risks or violations in the interconnected system before their occurrence. We propose to use a preset threshold to distinguish the SSR risks. For example, $\bar{M}_{\text{ssr}} > 0.2$ means safe and otherwise risky.

### 4.3 Preventive Control for SSR Mitigation

This subsection will develop a preventive controller to proactively manage the workload in order to achieve a decent level of safety margins. Through proactive actions, the controller contributes to enhancing data centers' capacity of SSR mitigation.

The preventive control is realized by running the following optimization model:

$$\max_{P_{\text{load}}} \left( \bar{M}_{\text{ssr}}(P_{\text{load}}) - \beta (P_{\text{load}} - P_{\text{set}})^2 \right) \tag{19}$$

where $P_{\text{set}}$ is the predicted workloads (setpoints) for the decision moment; $\beta$ is a weighting factor to balance the two objectives of safety improvement and setpoint tracking.

This is a typical two-layered optimization with an inner minimization (due to $\bar{M}_{\text{ssr}}$) and an outer maximization. We apply nested Bayesian optimization to solve this problem. In the implementation, we also consider the feasible lower and upper bound for $P_{\text{load}}$, and the feasible modification of $P_{\text{load}} - P_{\text{set}}$ is assumed to be bounded.

## 5. CASE STUDY

The case study builds upon a representative medium-sized AI data center that is equipped with a cluster of compute servers. These servers are almost homogeneous and responsible to serve the downstream compute workloads. For the configuration, a typical 3.6 kW server is designed to power eight A100 GPU (450 W per chip) through one PFC converter that transforms 230 V AC(rms) into 400 V DC supplies. The detailed circuit parameters are provided in the supplementary material, which are consistent with the current industrial practices and are carefully tuned for decent performances.

The workload is assumed to be an adjustable constant setpoint, and the supervisory control system may automatically limit or delay the workload when necessary. In the proposed cooperative framework, this data center will be updated with the grid information and can further compute the Thevenin equivalence outside a PFC converter. For brevity, we take account of such a representative testbed with a PFC connecting the grid side and serving a constant power load. All the circuit simulations are validated using LTspice (a SPICE-based simulator). The data-driven models and algorithms are realized in Python and conducted on a Macbook Pro (M3 chip).

We typically focus on the following three questions:

- What is unique characteristic of PFC impedance, and in what ways does it depend on workload?
- How to efficiently evaluate the safety status of a given operating condition? How should we issue early warnings and what message will be informative?
- Is deloading always a robust strategy for safety improvement? What is the best way to achieve safety benefits with minimal workload adjustments?

## 5.1 Impedance Across Different Workloads

This subsection will dive into the detailed impedance features and demonstrate the workload dependence.

It is found that PFC converters exhibit special impedance characteristics with discernible differences with respect to workload conditions. *Fig. 6* visualizes the impedance $Z_{\text{gcdc}}$ of two workload cases using the Bode plots and the Nyquist plots.

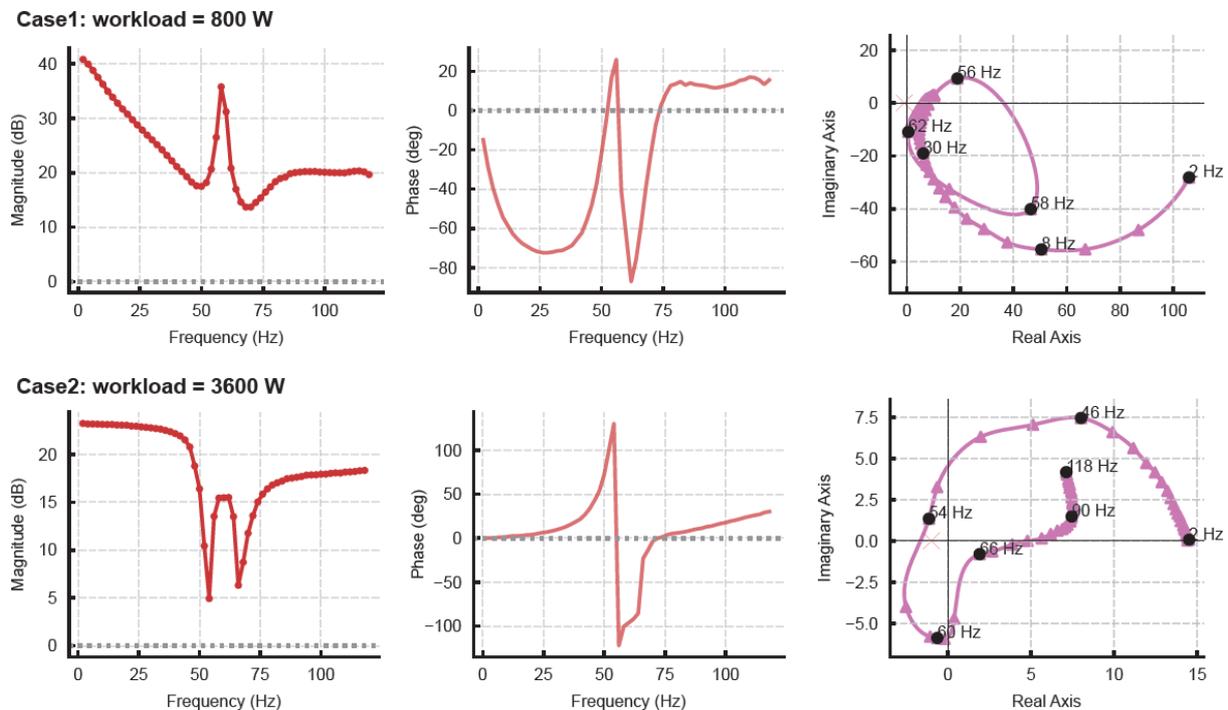

*Fig. 6. Bode plots and Nyquist plots of converter impedance for two workload cases.*

One can find that low workloads contribute to an increase in impedance magnitudes but the phase response is limited below 25 degrees. The associated complex-valued trajectory is mostly located at the fourth quadrant of the complex plane. On the contrary, high workloads have shown a more fluctuating phase response with the trajectory traversing different quadrants as frequency increases.

Despite the differences, both workload cases exhibit two dips in their magnitude responses around 60 Hz. This is a unique and inherent characteristic of PFC converters. Remind that SSR conditions require an equality between $\|Z_{\text{grid}}\|$ and $\|Z_{\text{gcdc}}\|$, these magnitude dips appear as the potential weaknesses that may trigger resonance incidents.

To further examine the diversity, we construct impedance vectors by directly concatenating the real/imaginary (rectangular form) or magnitude/phase values (phasor form) at all frequency measurements. A hierarchical clustering is then deployed to get two dendrograms in *Fig. 7*.

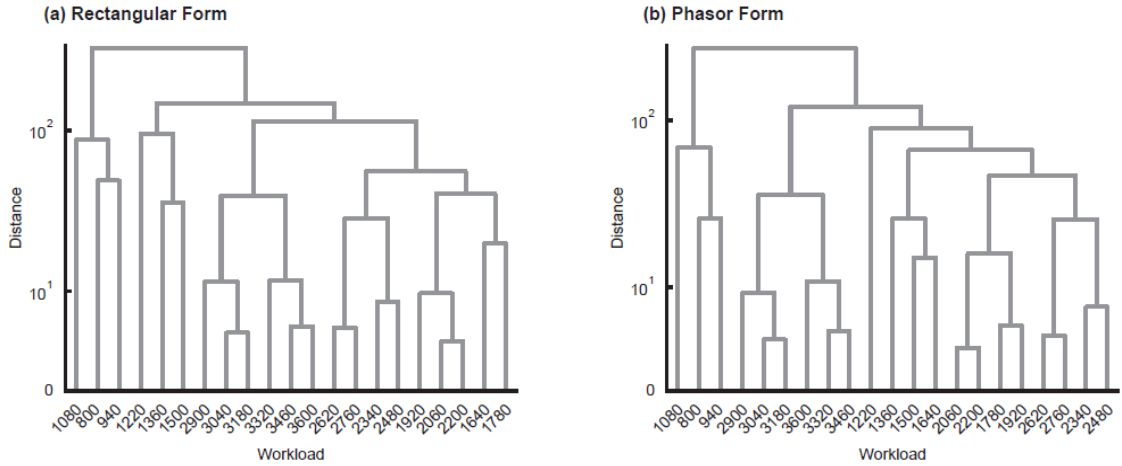

*Fig. 7. Dendrograms of the hierarchical clustering on impedance.*

Both cluster hierarchies agree to separate the low workloads (800~1040 W) as a single group, indicating that idle-like workload seem to be very different from the rest. Actually in the circuit simulation, low workloads are likely to cause duty cycles frequently touching the allowed lower bound. These clustering results also provide strong evidence of the heterogeneity of impedance characteristics, which further requires complicated, fine-grained modeling for SSR mitigation.

Next, we will apply machine learning to accelerate the estimation process of impedances. It turns out that the learning-based estimation not only bypass the expensive circuit simulations (5~15 minutes, up to hours), but also performs very impressive estimation accuracy. *Tab. 1* summaries the accuracy and running time of different model options.

*Tab. 1. Estimation performance of different machine learning models*

| Method | MSE | sMAPE | Training Time (s) | Inference Time (ms) |
| --- | --- | --- | --- | --- |
| SVM | 4.342 | 0.203 | 28.174 | 8.272 |
| Adaboost | 1.06 | 0.214 | 0.715 | 43.456 |
| Dense Net (5 layers) | 1.531 | 0.243 | 64.344 | 5.651 |
| Dense Net (6 layers) | 0.783 | 0.202 | 68.942 | 2.362 |

*Note: the hidden layers of 5-layer neural nets and 6-layer neural nets are 16-32-16, 16-16-16-16 accordingly.*

The mean-squared errors and symmetric mean absolute percentage errors are adopted to measure the accuracy (smaller is better). According to these two metrics, the dense neural net of six layers outperform all other options. An obvious benefit of neural nets is the fast inference, and this will finally enable large-scale online computation.

## 5.2 Performance of Early Warnings and Preventive Control

This subsection will contextualize the early warning of SSR risks, and then analyze the decisions of preventive control for SSR mitigation.

Efforts are made to demonstrate that the safety margin is essentially a distance measure in the complex plane. *Fig. 8* illustrates this idea with a concrete example. Here, the trajectory encircles the critical point of $-1 + j0$ and the closest distance is marked green and labeled as safety margin in the zoom-in view.

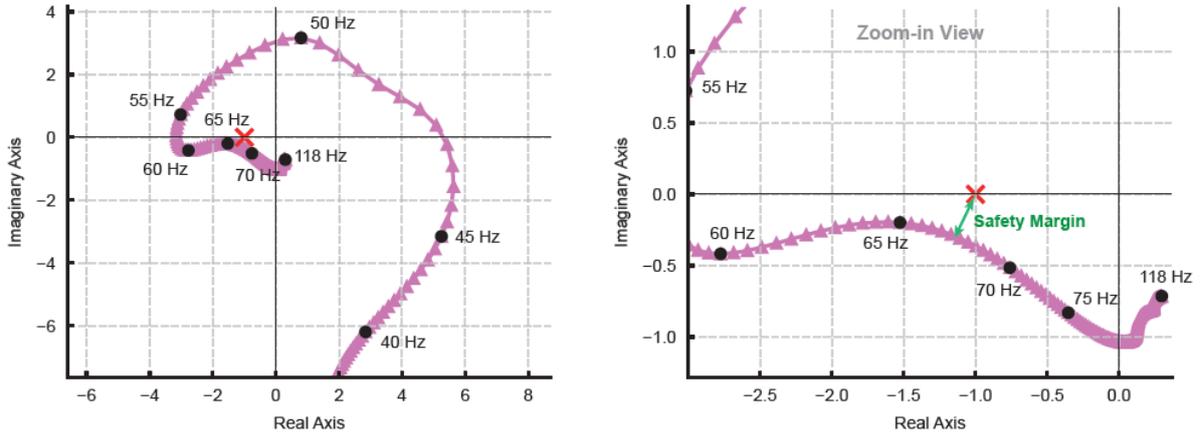

*Fig. 8. Safety margin demonstration.*

Early warning is a monitoring process that takes a quick and accurate estimation when the safety margin may fall below a threshold value. It also supports full preparation ahead of time for diverse selective scenarios. *Tab. 2* presents the normalized safety margins that are simulated under different combinations of grid conditions and workload setpoints.

*Tab. 2. Safety margins under diverse operating conditions.*

| Conditions | 800W | 2060W | 2480W | 3040W | 3460W |
|---|---|---|---|---|---|
| **12mH** | 0.095 | 0.243 | 0.172 | 0.157 | 0.277 |
| **7mH** | 0.103 | 0.271 | 0.229 | 0.262 | 0.402 |
| **2mH** | 0.107 | 0.286 | 0.286 | 0.339 | 0.457 |

*Note: each column represents different workloads, and each row is different in the grid impedance.*

Two clear messages from *Tab. 2* include: higher workloads generally have better safety margins; and stiff grid conditions (smaller grid inductance) increase the stability of interconnection. If setting 0.2 as the threshold, five conditions from *Tab. 2* should be detected as abnormal and mitigation strategies may apply accordingly. In early warning reports, the safety margins for a certain level of grid impedance fluctuations could be a useful information to guide robust solutions.

A full scan over workload levels is done in *Fig. 9* to illustrate how workload may affect the safety margins. As workload is controllable in data centers, the findings from this simulation will motivate and inform our design of preventive controllers.

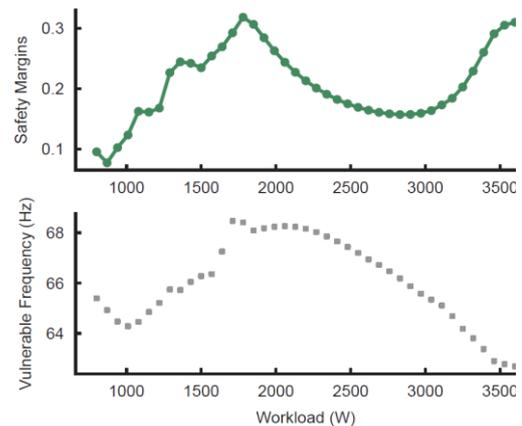

It clearly shows that there are two disjoint regions of high safety margins, i.e., 1290~2270 W and 3250~3600 W. A basin is found between them, which can drop down to half of the peak margins at 1780 W. This finding is unexpected and against the conventional belief in deloading strategies. The nonlinearity shown in *Fig. 9* complicates our decisions in workload management, because both decreasing and increasing workload setpoints will contribute to moving out of the basin.

*Fig. 9. Workload impacts on safety margins.*

We focus on the control performance and showcase the inherent trade-off between safety improvement and minimal workload rescheduling. *Tab. 3* summarizes the control performances and decision details under different conditions.

*Tab. 3. Optimal workload management under different conditions*

| Lgrid (mH) | Pset (W) | Pload (W) | Wvul (Hz) | Margin Increase | Power Diff (W) |
|---|---|---|---|---|---|
| 12 | 2060 | 1792.29 | 68.111 | 0.077 | -267.71 |
| 12 | 3040 | 3522.299 | 62.706 | 0.133 | 482.299 |
| 7 | 2060 | 1851.605 | 70.465 | 0.040 | -208.395 |
| 7 | 3040 | 3582.997 | 64.203 | 0.180 | 542.997 |

Diverse decisions are observed in the first two and the last two rows, which reflects the inherent nonlinearity and non-convexity. The preventive controller suggests deloading when the power setpoint is 2060 W, but the decision flips for a higher power setpoint of 3040 W. Multiple evaluations are often required to accurately resolve the trade-offs and finally achieve safety benefits with minimal workload adjustments.

## 6. CONCLUSION

This paper is motivated to investigate the rich interactions between data centers and the electric grid, which may pose additional risks that are largely overlooked at the current practices. We further propose a cooperative mechanism for data center managers and grid operators to achieve the goal of enhanced system reliability.

Technically, we focus on grid-connected data centers and justify why data centers may trigger SSR incidents from the perspective of impedance analysis. Through high-precision impedance modeling and frequency sweep, this paper demonstrates the unique characteristics of power factor correction converters and showcases the dependence on compute workloads. A learning-based model is established as a surrogate for fast online impedance estimation. The distance-based stability metric is proposed to support the design of a preventive controller that executes workload management to mitigate SSR risks. This controller is simple-yet-effective and it will secure the future electric grid with massive deployment of data centers.

Overall, the technical insights from this paper are expected to inform the risk management and resonance mitigation for grid operators as well as investment guidelines for future data center projects.